\documentclass[12pt]{iopart}
\usepackage{graphicx, iopams}
\usepackage{hyperref}

\begin{document}

\title[]{Jacobi Crossover Ensembles of Random Matrices and Statistics of Transmission Eigenvalues}

\author{Santosh Kumar and Akhilesh Pandey}

\address{School of Physical Sciences, Jawaharlal Nehru University, 
New Delhi - 110067, India}
\eads{\mailto{skumar.physics@gmail.com}, \mailto{ap0700@mail.jnu.ac.in}}
\begin{abstract}
We study the transition in conductance properties of chaotic mesoscopic cavities as time-reversal symmetry is broken. We consider the Brownian motion model for transmission eigenvalues for both types of transitions, viz., orthogonal-unitary and symplectic-unitary crossovers depending on the presence or absence of spin-rotation symmetry of the electron. In both cases the crossover is governed by a Brownian motion parameter $\tau$, which measures the extent of time-reversal symmetry breaking. It is shown that the results obtained correspond to the Jacobi crossover ensembles of random matrices. We derive the level density and the correlation functions of higher orders for the transmission eigenvalues. We also obtain the exact expressions for the average conductance, average shot-noise power and variance of conductance, as functions of $\tau$, for arbitrary number of modes (channels) in the two leads connected to the cavity. Moreover, we give the asymptotic result for the variance of shot-noise power for both the crossovers, the exact results being too long. In the $\tau \rightarrow 0$ and $\tau \rightarrow \infty$ limits the known results for the orthogonal (or symplectic) and unitary ensembles are reproduced. In the weak time-reversal symmetry breaking regime our results are shown to be in agreement with the semiclassical predictions.
\end{abstract}

\pacs{73.23.-b, 73.63.Kv, 72.15.Rn, 05.45.Mt}

\maketitle

\section{Introduction}
The transport of electrons through ballistic mesoscopic cavities has emerged as a subject of intensive research in recent years \cite{MK,KMM,Bnkr,YA,HPM,OSS,BB,Rich,Eftv,BM,JPB,LW,Br,Lndr,FL,Btkr,BBtkr,SS,SSW,KFP1,KFP2,Zuk,PW,FSS,BHMH,BC,Nov,BAB,ABSW,AM,VV,For}. The technological advancement has made it possible to fabricate mesoscopic cavities of arbitrary geometries and degree of disorder \cite{MK, KMM, Bnkr, YA, HPM, OSS}. The main theoretical tools used to  study the quantum conductance phenomenon in these systems include random matrix theory (RMT) \cite{Bnkr}, diagrammatic methods \cite{BB}, semiclassical methods \cite{Rich} and supersymmetry techniques \cite{Eftv}. RMT provides a convenient way to study the transport properties when the underlying classical dynamics of these cavities is chaotic \cite{MK, KMM, Bnkr, YA, BM, JPB}, and it has been quite successful.

The scattering matrix $S$, which relates the electron fluxes in the two leads connected to the mesoscopic cavity, is central to the problem of quantum transport. Under RMT the scattering matrix is assumed to belong to one of Dyson's standard circular ensembles. A microscopic justification for using circular ensembles as models for scattering matrices has been provided by calculations involving the Gaussian ensembles \cite{LW,Br}. The appropriate circular random matrix ensemble for the problem is decided by the time-reversal and rotational symmetries of the system and is determined by its invariance under orthogonal, symplectic or unitary transformations \cite{Bnkr,RM}. The invariant ensembles are abbreviated as OE, SE and UE respectively and characterize the three classes of chaotic cavities. 

Various physical quantities of interest such as average conductance, average shot-noise power and variance of conductance derive from the statistics of transmission eigenvalues, which are obtained by considering the polar decomposition of the scattering matrix. It turns out that the statistics of transmission eigenvalues $T_j$ for the invariant cases is related to the statistics of eigenvalues of a special case of Jacobi ensembles under a simple linear transformation of $T_j$ \cite{Bnkr,AM,VV,For}. The relation between transmission eigenvalues obtained from the scattering matrix and the eigenvalues of Jacobi ensembles of random matrices has been explicitly shown in \cite{For} for the invariant cases. 

The problem of statistics intermediate between above invariant classes is also important. For instance, application of a weak magnetic field to the system breaks the time-reversal symmetry partially. In this case, depending on the presence or absence of spin-rotation symmetry of the electron, statistics intermediate between OE and UE or SE and UE is observed. The problem of intermediate statistics in OE-UE and SE-UE transitions has been solved exactly for the energy level correlations as modeled by the Gaussian \cite{PM,MP} and circular \cite{PS} ensembles and recently for the entire Jacobi family of random matrix ensembles \cite{SKP}.  

Our purpose in this paper is to study the intermediate statistics for the conductance in mesoscopic systems. This has been a subject of interest both theoretically and experimentally \cite{KFP1,KFP2,Zuk,PW,FSS,BHMH,BC,Nov,BAB,ABSW}. The problem of intermediate statistics is studied under the framework of Dyson's Brownian motion model of random matrices \cite{PS,Dys62}. The corresponding Brownian motion model for transmission eigenvalues has been obtained by Frahm and Pichard \cite{KFP1} using the polar decomposition of the scattering matrices. This problem of finding the transmission eigenvalue statistics for the crossovers is technically challenging, because there is no direct relation between the transmission eigenvalues and the eigenphases of scattering matrix \cite{PS}. Frahm and Pichard have considered the model for equal number of incoming and outgoing channels and solved the OE-UE crossover problem \cite{KFP2}. We extend their formalism to unequal number of incoming and outgoing channels for both OE-UE and SE-UE crossovers. We show that the results for the transmission correlation functions correspond to Jacobi crossover ensembles \cite{SKP}. We use these results to calculate the above-mentioned physical observables as functions of the Brownian motion parameter $\tau$. In the $\tau\rightarrow0$ and $\tau\rightarrow\infty$ limits, results corresponding to OE (or SE) and UE, are reproduced. Also in the weak time-reversal symmetry breaking regime our results are consistent with the semiclassical predictions \cite{BHMH, BC}.

The paper is organized as follows. In section \ref{secBM} we make use of the Brownian motion model to obtain the jpd of transmission eigenvalues in the crossover regimes. In section \ref{secCC} we introduce the two-point kernels which are used to calculate the $n$-level correlation functions. Also, we express the averages and variances of conductance and shot-noise power in terms of moments involving these kernels. In section \ref{secOU} we consider the OE-UE crossover and give the explicit expression for the density of transmission eigenvalues. This generalizes the result available for the UE \cite{AM, VV} to the crossover. Using the Landauer-B\"{u}ttiker formalism \cite{Lndr,FL,Btkr,BBtkr} we find the exact results for the above mentioned physical quantities for the transition.  Section \ref{secSU} consists of similar results for the SE-UE crossover. In section \ref{secSC} we compare our results with semiclassical predictions in the weak symmetry breaking regime. Section \ref{secUCF} deals with the phenomenon of universal fluctuations, where we compare the variances of conductance and shot-noise power for the OE-UE and SE-UE crossovers. We conclude in section \ref{secC} with some general discussion. Derivation of the skew-orthogonal polynomials and the associated two-point kernels are given in the appendices.

%~~~~~~~~~~~~~~~~~~~~~ Section II ~~~~~~~~~~~~~~~~~~~~~~~~
\section{The Brownian Motion Model}
\label{secBM}

In this section we first consider transitions to all three invariant ensembles OE, UE, SE. We use the parameter $\beta$ with values 1, 2, 4 respectively for the three cases. In the later part of this section as well as in the rest of the paper we specialize to the case of $\beta=2$.

The scattering matrix $S$ for the conductance problem has the standard decomposition in terms of reflection matrices $r,r'$ and transmission matrices $t,t'$ as \cite{MK, Bnkr, For}
\begin{equation}
\label{smat}
S=\left(\begin{array}{cc}
  r_{N_1\times N_1} & t'_{N_1\times N_2} \\
  t_{N_2\times N_1} & r'_{N_2\times N_2}
  \end{array}\right).
\end{equation}
Here $N_1$ and $N_2$ refer to the number of modes in the left and right leads connected to the cavity. So the matrices $r_{N_1\times N_1}$ and $r'_{N_2\times N_2}$ correspond to the reflection from left to left and right to right respectively. Similarly $t_{N_2\times N_1}$ and $t'_{N_1\times N_2}$ represent the transmission from left to right and right to left respectively. The scattering matrix $S$ is thus $N_\mathrm{s}=N_1+N_2$ dimensional. Also we write $N=\mbox{min}(N_1,N_2)$. As a consequence of unitarity of $S$, the Hermitian matrices $t^\dag t, t'^\dag t', 1-r^\dag r$ and $1-r'^\dag r'$ have $N$ common eigenvalues $T_1,...,T_N$ with values between $0$ and $1$. We denote the set of these transmission eigenvalues as $\{T\}$.

For considering the transition problem it is assumed that the evolution of $S$ takes place as
\begin{equation}
\label{sevol}
S(\tau+\delta\tau)=S(\tau)e^{i\sqrt{\delta\tau}M(\tau)},
\end{equation}
where $M(\tau)$, independent for each $\tau$, belongs to one of the invariant classes of Gaussian ensembles of random matrices \cite{PS,Dys62}. It can be shown \cite{KFP1} that the jpd of transmission eigenvalues $\mathcal{P}(\{T\};\tau)$ obeys the following Fokker-Planck equation: 
\begin{equation}
\label{FP}
\frac{\partial \mathcal{P}(\{T\};\tau)}{\partial \tau}=-\mathcal{L}\mathcal{P}(\{T\};\tau).
\end{equation}
In the above equation the Fokker-Planck operator $\mathcal{L}$ is given by
\begin{equation}
\label{FPop}
\mathcal{L}=\sum_{j=1}^N\frac{\partial}{\partial T_j}e^{-\beta W}\frac{\partial}{\partial T_j}T_j(1-T_j) e^{\beta W},
\end{equation}
with
\begin{equation}
\label{work}
W=-\mbox{ln}\Big[|\Delta_N(\{T\})|\prod_{k=1}^N T_k^{(|N_1-N_2|+1)/2}(1-T_k)^{1/\beta}\Big].
\end{equation}
Here $\Delta_N(\{T\})=\prod_{j<k}(T_j-T_k)$ is the Vandermonde determinant. The equilibrium jpd is obtained from $\mathcal{L}\mathcal{P}_{\mathrm{eq}}^{(\beta)}=0$ and is given by \cite{Bnkr,For}
\begin{equation}
\label{PeqT}
\mathcal{P}_{\mathrm{eq}}^{(\beta)}(\{T\})=\mathcal{C}_N^{(\beta)} |\Delta_N(\{T\})|^\beta \prod_{j=1}^N T_j^{\frac{\beta}{2}(|N_1-N_2|+1-\frac{2}{\beta})}.
\end{equation}                                                                       
To make contact with Jacobi weight function we make the transformation 
\begin{equation}
\label{x-T}
x_j=2T_j-1,
\end{equation}
and then $0\leq T_j \leq 1$ implies $-1 \leq x_j \leq 1$. Thus we get the jpd $P_{\mathrm{eq}}^{(\beta)}(\{x\})$ for the $x$ variables, 
\begin{equation}
\label{PeqX}
P_{\mathrm{eq}}^{(\beta)}(\{x\})=C_N^{(\beta)} |\Delta_N(\{x\})|^\beta \prod_{j=1}^N (1+x_j)^{\beta(b+1-\frac{1}{\beta})},
\end{equation}
where we have introduced 
\begin{equation}
2b+1=|N_1-N_2|. 
\end{equation}
 The Jacobi weight function involved in the above jpd is now easily recognized as $w_{\mu,\nu}(x)=(1-x)^\mu(1+x)^\nu$, with $\mu=0$ and $\nu=\beta(b+1-1/\beta)$. In appendix A, we give the relevant results for the Jacobi polynomials. The similarity transformation $\xi=(P_{\mathrm{eq}}^{(\beta)})^{-1/2}P$ in the Fokker-Planck equation leads to the following Schr\"{o}dinger equation for $\xi(\{x\};\tau)$ in imaginary time $i\tau$:
\begin{equation}
\label{Schro}
\frac{\partial\xi(\{x\};\tau)}{\partial \tau}=-\mathcal{H}\xi(\{x\};\tau),
\end{equation} 
where 
\begin{equation}
 \mathcal{H}=(P_{\mathrm{eq}}^{(\beta)})^{-1/2}\mathcal{L}\,(P_{\mathrm{eq}}^{(\beta)})^{1/2}
\end{equation}
is Calogero-Sutherland type of Hamiltonian. Here $\mathcal{L}$ is given by (\ref{FPop}), (\ref{x-T}) in the $x_j$ variables.

For transitions to UE we have $\beta=2$, and this leads to the Jacobi weight function
\begin{equation}
w_{0,2b+1}(x)=(1+x)^{2b+1}
\end{equation}
in (\ref{PeqX}). Moreover in this case the Hamiltonian $\mathcal{H}$ splits into $N$ single-fermion Hamiltonians minus the ground state energy of the $N$ fermions described by $\mathcal{H}_x$, viz. $\mathcal{H}=\sum_{j=1}^N \mathcal{H}_{x_j}-\mathcal{E}_{0}$. Here
\begin{eqnarray}
\label{Hx}
\fl
\mathcal{H}_x\!\!&=&\!\!-\bigg[(1-x^2)\frac{\partial^2}{\partial x^2}-2x\frac{\partial}{\partial x}-\left(b+\frac{1}{2}\right)^2\left(\frac{1-x}{1+x}\right)+\left(b+\frac{1}{2}\right)\bigg].
\end{eqnarray}
The eigenfunctions of $\mathcal{H}_x$ are the weighted Jacobi Polynomials $w_{0,b+1/2}(x)P_n^{0,2b+1}(x)$ with eigenvalues
\begin{equation}
\label{En}
\boldsymbol{\varepsilon}_n=n(n+2b+2).
\end{equation}
Also $\mathcal{E}_{0}=\sum_{n=0}^{N-1}\boldsymbol{\varepsilon}_n$, so that the ground state energy of $\mathcal{H}$ is zero. 

The formal solution of (\ref{FP}) in terms of the $x_j$ variables is given by
\begin{equation}
P(\{x\};\tau)=e^{-\mathcal{L}\tau}P(\{x\};0).
\end{equation}
Then using the above results, the jpd of eigenvalues for OE-UE and SE-UE transitions can be written as \cite{SKP}
\begin{equation}
\label{Ptrans}
P(\{x\};\tau)=C_N\Delta_N(\{x\})\mbox{Pf}[F_{j,k}^{(\tau)}]\prod_{i=1}^N w(x_i),
\end{equation}
where $C_N$ is the normalization, Pf represents Pfaffian and $w(x)$ is the initial weight function. In sections \ref{secOU} and \ref{secSU} we choose $w(x)$, such that the initial jpd corresponds to (\ref{PeqX}) with $\beta=1$ and $4$ respectively. $F_{j,k}^{(\tau)}$ is an antisymmetric function which can be obtained from its $\tau=0$ counterpart using the one-body operators introduced in \cite{SKP}; see appendix B. The indices $j,k$ take the values from 1 to $N$ or $N+1$ depending on whether $N$ is even or odd respectively. When $N$ is even $F_{j,k}^{(0)}$ is simply $\mbox{sgn}(x_j-x_k)/2$ and $-\delta'(x_j-x_k)$ respectively for the orthogonal and symplectic cases. For odd $N$ one has to introduce additional term $F_{j,N+1}^{(\tau)}$ for OE-UE crossover as discussed in appendix B. In the SE Kramers degeneracy requires the levels to be doubly degenerate; thus we choose both $N_1$ and $N_2$ to be even so that $N$ is always even.

%~~~~~~~~~~~~~~~~~~~~~ Section III ~~~~~~~~~~~~~~~~~~~~~~~~
\section{Correlation Functions and Conductance Properties} 
\label{secCC}

For the calculation of $n$-level correlation function $R_n(x_1,...,x_n;\tau)$, given by
\begin{equation}
\label{Rn}
R_n(x_1,...,x_n;\tau)=\frac{N!}{(N-n)!}\int_{-1}^{1}\cdots \int_{-1}^{1}P(\{x\};\tau)dx_{n+1}...dx_{N},
\end{equation}
we need the two-point kernels $S_N^{(\tau)}(x,y)$, $A_N^{(\tau)}(x,y)$ and $B_N^{(\tau)}(x,y)$, which in turn depend on the weighted skew-orthogonal polynomials $\phi_j^{(\tau)}(x)$ and the integrated (dual) functions $\psi_j^{(\tau)}(x)$ \cite{SKP}. These can again be obtained from their initial counterparts ($\tau=0$ results) \cite{PG2000,GP}. See appendix C for the skew-orthogonal polynomials and appendix D for the kernels. The polynomials are worked out explicitly in appendices E and F respectively for OE-UE and SE-UE transitions, and the corresponding kernels are given in appendices G and H. 

The skew orthogonality relation and application of Dyson's theorems \cite{RM,Dys70} lead to the quaternion determinantal expression for $R_n$ involving the above mentioned two-point kernels; see (\ref{Qdet}).
The $n$-level correlation function $\mathcal{R}_n$ for the transmission eigenvalues is related to $R_n$ as
\begin{equation}
\label{RTRx}
 \mathcal{R}_n(T_1,...,T_n;\tau)=2^n R_n\left(2T_1-1,...,2T_n-1;\tau\right).
\end{equation}

We now consider the application of these results to the quantum conductance problem and find expressions for the quantities of interest, viz. the averages and variances of both conductance and shot-noise power. The calculation of these quantities involve only the level density $R_1(x;\tau)$ and the two-level correlation function $R_2(x,y;\tau)$. These are given in terms of the above kernels as
\begin{equation}
\label{R1}
 R_1(x;\tau)=S_N^{(\tau)}(x,x)
\end{equation}
and
\begin{eqnarray}
\label{R2}
\fl
R_2(x,y;\tau)\!\!&=&\!\!R_1(x;\tau)R_1(y;\tau)-S_N^{(\tau)}(x,y)S_N^{(\tau)}(y,x)+A_N^{(\tau)}(x,y)B_N^{(\tau)}(x,y). 
\end{eqnarray} 
We also need the following moments of $R_1$ and $R_2$,
\begin{equation}
\label{mx}
\left<x\right>=\int_{-1}^{1}x R_1(x;\tau)dx,
\end{equation}
\begin{equation}
\label{mx2}
\left<x^2\right>=\int_{-1}^{1}x^2 R_1(x;\tau)dx,
\end{equation}
\begin{equation}
\label{mxy}
\left<xy\right>=\int_{-1}^{1}\int_{-1}^{1}xy R_2(x,y;\tau)dx\,dy,
\end{equation}
along with the normalizations
\begin{equation}
\int_{-1}^{1}R_1(x;\tau)dx=N,
\end{equation}
\begin{equation}
\int_{-1}^{1}\int_{-1}^{1}R_2(x,y;\tau)dx\,dy=N(N-1). 
\end{equation}
These moments along with similar averages $\left<x^4\right>$ and $\left<x^2 y^2\right>$ (needed for the variance of shot-noise power) can be calculated using $R_1(x;\tau)$, $R_2(x,y;\tau)$ given by (\ref{R1}), (\ref{R2}) and the kernels given in appendices G and H. This requires some tedious algebra involving repeated use of the orthogonality and the three-term recurrence relations for the Jacobi polynomials as given in appendix A. We use these quantities in sections \ref{secOU} and \ref{secSU} for OE-UE and SE-UE crossovers.
 
Landauer-B\"{u}ttiker formalism \cite{MK,Bnkr,Lndr,FL,Btkr,BBtkr} enables us to determine a variety of transport properties of the mesoscopic cavities using the statistics of transmission eigenvalues. Of particular interest are the observables which are linear statistics on the transmission eigenvalues, such as conductance and shot noise. These are quantities which do not contain products of different eigenvalues. We start with the dimensionless conductance which, at zero temperature, is given by the Landauer formula \cite{Bnkr,Lndr,FL}
\begin{equation}
g=\sum_{j=1}^N T_j.
\end{equation}
Thus it follows that the average conductance is given by
\begin{equation}
\label{Gx}
\left<g\right>=\int_0^1 T\; \mathcal{R}_1(T;\tau)dT=\frac{N}{2}+\frac{\left<x\right>}{2}.
\end{equation}
Next we consider the shot noise. These are the time-dependent fluctuations caused due to the quantum nature of electron charge, and persist down to zero temperature. The expression for the shot-noise power (dimensionless) is known due to B\"uttiker \cite{Bnkr,Btkr,BBtkr} as
\begin{equation}
p=\sum_{j=1}^N T_j(1-T_j).
\end{equation}
From this we obtain the average shot-noise power
\begin{equation}
\label{Px}
\left<p\right>=\int_0^1 T(1-T)\; \mathcal{R}_1(T;\tau)dT=\frac{N}{4}-\frac{\left<x^2\right>}{4}.
\end{equation}
The variance of conductance is given by 
\begin{eqnarray}
\label{varGxy}
\nonumber
\mbox{var}(g)&=&\int_0^1 T_1^2\mathcal{R}_1(T_1;\tau)dT_1-(\int_0^1 T_1\mathcal{R}_1(T_1;\tau)dT_1)^2\\
\nonumber
&+&\int_0^1\int_0^1 T_1T_2\mathcal{R}_2(T_1,T_2;\tau)dT_1dT_2\\
&=&\frac{\left<x_1^2\right>}{4}-\frac{\left<x_1\right>^2}{4}+\frac{\left<x_1 x_2\right>}{4}.
\end{eqnarray}
Finally we have the variance of shot-noise power,
\begin{eqnarray}
\label{varPxy}
\nonumber
\mbox{var}(p)&=&\int_0^1 T_1^2(1-T_1^2)\mathcal{R}_1(T_1;\tau)dT_1-(\int_0^1 T_1(1-T_1)\mathcal{R}_1(T_1;\tau)dT_1)^2\\
\nonumber
&+&\int_0^1\int_0^1 T_1(1-T_1)T_2(1-T_2)\mathcal{R}_2(T_1,T_2;\tau)dT_1dT_2\\
&=&\frac{\left<x_1^4\right>}{16}-\frac{\left<x_1^2\right>^2}{16}+\frac{\left<x_1^2 x_2^2\right>}{16}.
\end{eqnarray}

The correlation functions for OE, UE and SE have been given explicitly in terms of Jacobi polynomials in \cite{PG2000,GP}. Using these one can work out the above mentioned averages which have been obtained earlier by other methods \cite{MK,Bnkr,SS,BHMH}. We find
\begin{equation}
\label{G_beta}
\left<g\right>=\frac{N_1 N_2}{N_\mathrm{s}-1+\frac{2}{\beta}},
\end{equation} 
\begin{equation}
\label{P_beta}
\left<p\right>=\frac{N_1N_2(N_1-1+\frac{2}{\beta})(N_2-1+\frac{2}{\beta})}{(N_\mathrm{s}-2+\frac{2}{\beta})(N_\mathrm{s}-1+\frac{2}{\beta})(N_\mathrm{s}-1+\frac{4}{\beta})},
\end{equation}

\begin{equation}
\label{VarG_beta}
\mbox{var}(g)=\frac{2N_1N_2(N_1-1+\frac{2}{\beta})(N_2-1+\frac{2}{\beta})}{\beta(N_\mathrm{s}-2+\frac{2}{\beta})(N_\mathrm{s}-1+\frac{2}{\beta})^2(N_\mathrm{s}-1+\frac{4}{\beta})}.
\end{equation}
As pointed out in \cite{SSW} we find that the exact expression for var($p$) does not have a compact form for $\beta=1,4$ for arbitrary $N_1,N_2$. For $\beta=2$ however the expression is somewhat simpler, given by
\begin{eqnarray}
 \label{VarP_U}
 \nonumber
\mbox{var}(p)&=&\frac{N_1^2 N_2^2 (N_1-1)^2 (N_2-1)^2}{(N_\mathrm{s}-3)(N_\mathrm{s}-2)^2(N_\mathrm{s}-1)^2(N_\mathrm{s})^2(N_\mathrm{s}+1)} \\
\nonumber
&+&\frac{N_1^2 N_2^2 (N_1+1)^2 (N_2+1)^2}{(N_\mathrm{s}-1)(N_\mathrm{s})^2(N_\mathrm{s}+1)^2(N_\mathrm{s}+2)^2(N_\mathrm{s}+3)}\\
&+&\frac{N_1^2 N_2^2 (N_1-N_2)^4}{(N_\mathrm{s}-2)^2(N_\mathrm{s}-1)(N_\mathrm{s})^2(N_\mathrm{s}+1)(N_\mathrm{s}+2)^2}.
\end{eqnarray}
Also when $N_1=N_2=N$, the exact expression for $\beta=1$ is given by 
\begin{equation}
\fl
 ~~~~~~\mbox{var}(p)=\frac{N(N+1)(8N^5+60N^4+142N^3+91N^2-49N-36)}{4(2N-1)(2N+1)^2(2N+3)^2(2N+5)(2N+7)},
\end{equation}
and for $\beta=4$ by
\begin{equation}
\fl
 ~~~~~~\mbox{var}(p)=\frac{N(2N-1)(128N^5-480N^4+568N^3-182N^2-49N+18)}{4(4N-7)(4N-5)(4N-3)^2(4N-1)^2(4N+1)}.
\end{equation}
For large $N_1, N_2$ we can write the compact $\beta$-dependent expression for var($p$) correct to $\mathcal{O}(1)$ as
\begin{equation}
\mbox{var}(p)=\frac{2}{\beta}\frac{N_1^2 N_2^2}{N_\mathrm{s}^8}[2N_1^2 N_2^2+(N_1-N_2)^4],
\end{equation}
which agrees with the result of \cite{SSW}.

We remark that these relations will also be obtained in sections \ref{secOU} and \ref{secSU} as $\tau\rightarrow 0$ and $\tau\rightarrow \infty$ limits of the transition results. Note moreover that the initial jpd for SE-UE transition contains explicitly doubly degenerate levels and therefore the above quantities should be suitably modified for $\beta=4$ when used in section \ref{secSU}.

%~~~~~~~~~~~~~~~~~~~~~ Section IV ~~~~~~~~~~~~~~~~~~~~~~~~
\section {OE-UE Crossover}
\label{secOU}

We consider here the OE-UE transition and calculate the above physical quantities as functions of Brownian motion parameter $\tau$. The initial jpd of eigenvalues (in terms of $x_j$) for this crossover (corresponding to OE) can be found by setting $\beta=1$ in (\ref{PeqX}). Thus the weight function appearing in (\ref{Ptrans}) in this case is $w(x)=w_{0,b}(x)$ and the crossover in the weight function is from $w_{0,b}(x)$ for OE to $w_{0,2b+1}(x)$ for UE. The jpd of eigenvalues for the transition can thus be read from (\ref{Ptrans}). Moreover all the two-point kernels can be worked out exactly giving thereby the correlation functions of all orders. See appendices D and G for explicit expressions. Using these one can work out $\left<x\right>$, $\left<x^2\right>$, $\left<x y\right>$ and other moments as mentioned earlier. The corresponding expressions for the transmission eigenvalues can be obtained by going back to the $T$ variables as in (\ref{x-T}) and (\ref{RTRx}).

\begin{figure*}[t]
\centering
\includegraphics*[width=0.98 \textwidth]{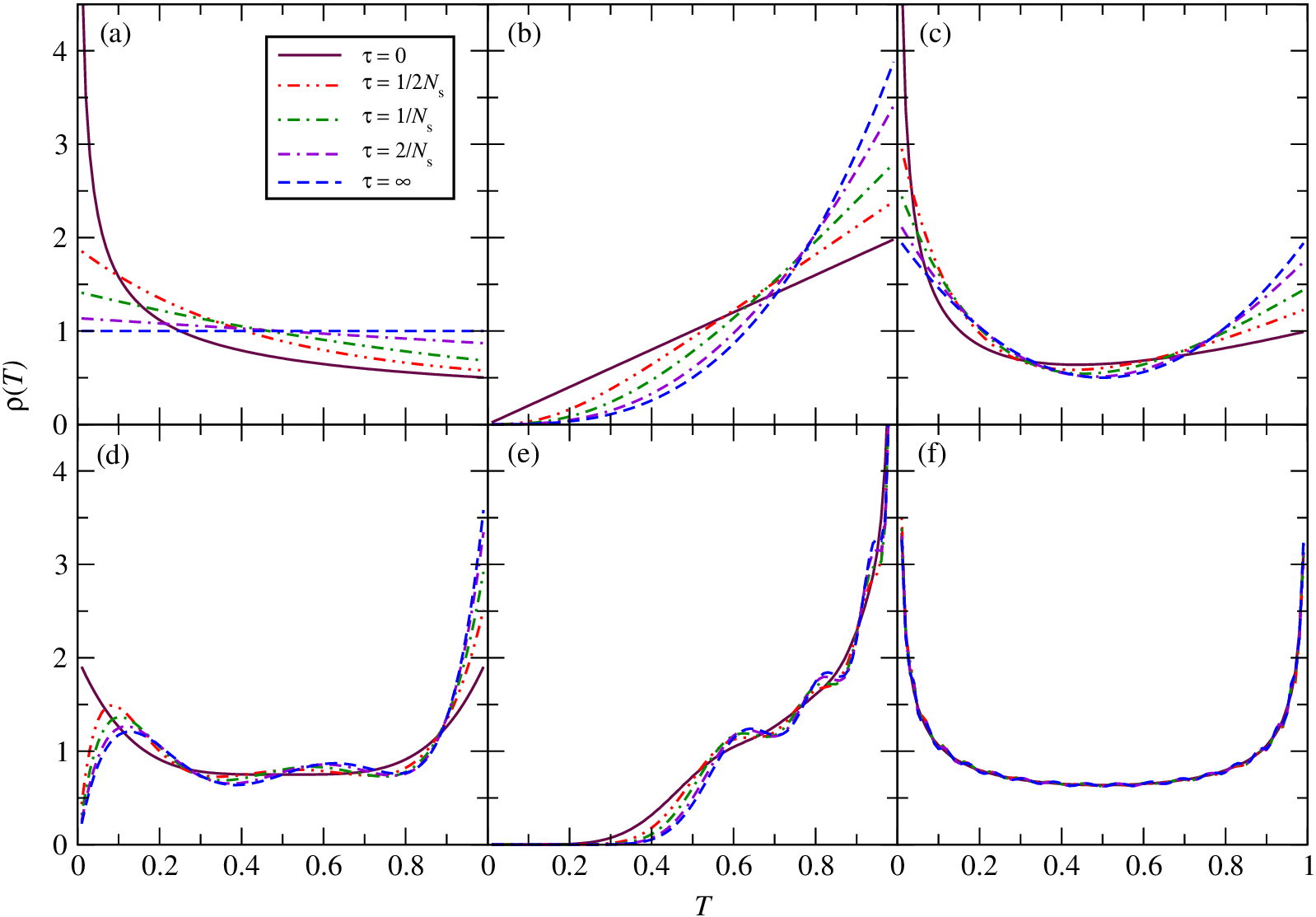}
\caption[]{The density of transmission eigenvalues $\rho(T)=\mathcal{R}_1(T)/N$ for OE-UE crossover for several values of $\tau$ for (a)$N_1=1,N_2=1$, (b)$N_1=1,N_2=4$, (c)$N_1=2,N_2=2$, (d)$N_1=3,N_2=4$, (e)$N_1=4,N_2=20$, (f)$N_1=20,N_2=20$.}
\end{figure*}
 
The level density $\mathcal{R}_1(T;\tau)$ (normalized to $N$) for transmission eigenvalues is found to be
\begin{eqnarray}
\label{Tdens_ou}
\fl
\mathcal{R}_1(T;\tau)=T^{2b+1}\sum_{\mu=0}^{N-1}(2\mu+2b+2)P_\mu^{0,2b+1}(2T-1)P_\mu^{0,2b+1}(2T-1)\nonumber\\
\fl ~~~~-T^{2b+1}\sum_{\mu=0}^{N-1}\sum_{\nu=N}^{\infty}(2\mu+2b+2)(-1)^{N+\nu}e^{-(\boldsymbol{\varepsilon}_\nu-\boldsymbol{\varepsilon}_\mu)\tau}P_\mu^{0,2b+1}(2T-1) P_{\nu}^{0,2b+1}(2T-1),
\end{eqnarray}
valid for both even and odd $N$. Here, on the RHS, the $\tau$-independent term, i.e. the first term, corresponds to the UE. The $\tau$-dependent term is the correction term which contributes in the non-equilibrium regime $(0\le\tau<\infty)$.
This term is what leads to the weak-localization correction in the average of quantities which form linear statistics on transmission eigenvalues. For $\tau\rightarrow\infty$, this term disappears completely and we retrieve the well known expression of level density for the UE \cite{AM,VV}. 

In the weak-localization study of relevant physical observables in mesoscopic systems the quantities are expressed in terms of diffuson and Cooperon contributions \cite{Zuk,PW,FSS}. These correspond respectively to the time-reversal symmetry independent and dependent parts. The $\tau$-independent and dependent terms in (\ref{Tdens_ou}) can be viewed as the analogues of the diffuson and Cooperon contributions.

For $N_1,N_2>>1$ and with fixed ratio $N_1/N_2$ the density is given for all $\tau$ by \cite{Bnkr}
\begin{equation}
\label{asymdens}
\mathcal{R}_1(T)=\cases{\frac{(N_1+N_2)}{2\pi T}\sqrt{\frac{T-T_c}{1-T}}, & $T_c\leq T\leq 1$,\\
~~~~~~~~0, & otherwise,\\}
\end{equation}

where
\begin{equation}
\label{Tc}
T_c=\left(\frac{N_1-N_2}{N_1+N_2}\right)^2. 
\end{equation}

The densities for various values of $N_1$ and $N_2$ are shown in figure 1 as functions of $\tau$. In figure 1(e) the densities vanish below some critical $T$ values which are somewhat smaller than $T_c\approx0.44$ predicted by (\ref{Tc}); this is because $N_1=4$ is not large enough. In figure 1(f) the densities for all $\tau$ are close to that given by (\ref{asymdens}) because $N_1=N_2=20$ is significantly large. 

The average conductance is obtained, by using the value of $\left<x\right>$ in (\ref{Gx}) along with the substitution $N=\mbox{min}(N_1,N_2)$ and $N+2b+1=\mbox{max}(N_1,N_2)$, as
\begin{eqnarray}
\label{Gt-O}
\left<g\right>=\frac{N_1 N_2}{N_\mathrm{s}}-\frac{N_1 N_2}{N_\mathrm{s}(N_\mathrm{s}+1)}e^{-N_\mathrm{s}\tau}.
\end{eqnarray}
Recall also that $N_{\mathrm{s}}=N_1+N_2$. Similarly, substituting the value of $\left<x^2\right>$ in (\ref{Px}) gives the average shot-noise power as
\begin{eqnarray}
\label{SNPt-O}
\fl
\nonumber
\left<p\right>=\frac{N_1^2N_2^2}{(N_\mathrm{s}-1)(N_\mathrm{s})(N_\mathrm{s}+1)}+\frac{N_1 N_2(N_1-N_2)^2}{(N_\mathrm{s}-2)(N_\mathrm{s})(N_\mathrm{s}+1)(N_\mathrm{s}+2)}e^{-N_\mathrm{s}\tau}\\
\fl
+\frac{N_1 N_2(N_1-1)(N_2-1)}{(N_\mathrm{s}-2)(N_\mathrm{s}-1)(N_\mathrm{s})(N_\mathrm{s}+1)}e^{-2(N_\mathrm{s}-1)\tau}-\frac{N_1 N_2(N_1+1)(N_2+1)}{N_\mathrm{s}(N_\mathrm{s}+1)(N_\mathrm{s}+2)(N_\mathrm{s}+3)}e^{-2(N_\mathrm{s}+1)\tau}.
\end{eqnarray}
The variance of conductance for OE-UE crossover is obtained from (\ref{varGxy}) as
\begin{eqnarray}
\label{varGt-O}
\fl
\nonumber
\mbox{var}(g)=\frac{N_1^2N_2^2}{(N_\mathrm{s}-1)(N_\mathrm{s})^2(N_\mathrm{s}+1)}+\frac{N_1 N_2(N_1-1)(N_2-1)}{(N_\mathrm{s}-2)(N_\mathrm{s}-1)(N_\mathrm{s})(N_\mathrm{s}+1)}e^{-2(N_\mathrm{s}-1)\tau}\\
\nonumber
-\frac{N_1^2 N_2^2}{(N_\mathrm{s})^2(N_\mathrm{s}+1)^2}e^{-2N_\mathrm{s}\tau}+\frac{N_1 N_2(N_1+1)(N_2+1)}{N_\mathrm{s}(N_\mathrm{s}+1)(N_\mathrm{s}+2)(N_\mathrm{s}+3)}e^{-2(N_\mathrm{s}+1)\tau}\\
+\frac{2N_1 N_2(N_1-N_2)^2}{(N_\mathrm{s}-2)(N_\mathrm{s})^2(N_\mathrm{s}+1)(N_\mathrm{s}+2)}e^{-N_\mathrm{s}\tau}.
\end{eqnarray}
The exact crossover result for the variance of shot-noise power can be obtained from (\ref{varPxy}) but it is too lengthy to be presented here. We give the asymptotic result in section \ref{secUCF}.

As expected, these results match with the results (\ref{G_beta})-(\ref{VarG_beta}) for $\beta=1,2$ in the limits $\tau\rightarrow 0$ and $\infty$ respectively. For $N_1=N_2$, (\ref{Gt-O}) and (\ref{varGt-O}) give back the corresponding results of Frahm and Pichard \cite{KFP2}. 

It is clear from the results (\ref{Gt-O})-(\ref{varGt-O}) that $N_\mathrm{s}\tau$ serves as the natural transition parameter for the problem. Thus for large $N_1, N_2$ non-trivial crossover is obtained when $\tau\sim1/N_\mathrm{s}$. The fact that the transition parameter scales as $\tau\sim 1/N_\mathrm{s}$ is not surprising. It is known, for example, in circular and Gaussian ensembles \cite{PS,Pandey} that the global properties such as the level density and variances of low-order traces scale in a similar way. This is in contrast to the local fluctuation properties, for example the spacing distribution and the number variance, where the transition occurs for $\tau\sim 1/N_\mathrm{s}^2$ \cite{PM,MP,PS,SKP,Pandey}. We also mention that the $\tau$-dependent terms give rise to $\mathcal{O}(1/N_\mathrm{s})$ corrections\footnote{The correction term itself is of $\mathcal{O}(1)$. By $\mathcal{O}(1/N_{\mathrm{s}})$ correction we mean $1/N_{\mathrm{s}}$ with respect to the leading term.} (the weak-localization corrections) in the respective leading terms of (\ref{Gt-O}) and (\ref{SNPt-O}). 

The symmetric case of $N_1=N_2>>1$ has been studied by many authors. The average conductance in this case is given by
\begin{equation}
\left<g\right>=\frac{N}{2}-\frac{1}{4}e^{-2N\tau},
\end{equation}
which yields the universal value $-1/4$ for the weak-localization correction in the limit $\tau\rightarrow 0$ \cite{Bnkr,BM,KFP2}.
Also in the expression of average shot-noise power given by (\ref{SNPt-O}), the first $\tau$-dependent term vanishes identically, whereas the last two terms cancel each other. As a result no weak-localization correction is obtained in the average shot-noise power. This is in agreement with the earlier predictions of absence of weak-localization correction in average shot-noise power in the symmetric case \cite{Bnkr,JPB}.

The variances of conductance and shot-noise power in the asymptotic limit lead to universal results and is considered in section \ref{secUCF}.

%~~~~~~~~~~~~~~~~~~~~~ Section V ~~~~~~~~~~~~~~~~~~~~~~~~
\section {SE-UE crossover}
\label{secSU}
In this section we consider the SE-UE crossover. Since we have to take into account Kramers degeneracy, the initial jpd cannot be obtained by direct substitution of $\beta=4$ in (\ref{PeqX}). Rather we need to consider $N_1\rightarrow N_1/2$ and $N_2\rightarrow N_2/2$ along with the introduction of $\delta$ functions to take care of degeneracy. Thus there are total of $N/2$ eigenvalues at $\tau=0$, each of them two-fold degenerate. As soon as $\tau\ne0$ the degeneracy is broken and we have $N$ distinct eigenvalues. It turns out that the initial weight function for this crossover is $w_{0,b+1}(x)$. The jpd of eigenvalues in this case is thus given by (\ref{Ptrans}) with $w(x)=w_{0,b+1}(x)$. Also the expressions for the two-point kernels can again be obtained exactly and then $R_n$ can be written down (see appendices D and H). 

We get the following expression for the level density $\mathcal{R}_1(T;\tau)$ for transmission eigenvalues in this case:
\begin{eqnarray}
\label{Tdens_su}
\fl
\mathcal{R}_1(T;\tau)=T^{2b+1}\sum_{\mu=0}^{N-1}(2\mu+2b+2)P_\mu^{0,2b+1}(2T-1)P_\mu^{0,2b+1}(2T-1)\nonumber\\
\fl
~~~ -T^{2b+1}\sum_{\mu=0}^{N-1}\sum_{\nu=N}^{\infty}(2\nu+2b+2)(-1)^{N+\mu}e^{-(\boldsymbol{\varepsilon}_\nu-\boldsymbol{\varepsilon}_\mu)\tau}P_\mu^{0,2b+1}(2T-1) P_{\nu}^{0,2b+1}(2T-1).
\end{eqnarray}
Again as in the OE-UE case the first term in the above equation corresponds to the UE and the second term is responsible for the weak-antilocalization corrections. The similarity between (\ref{Tdens_ou}) and (\ref{Tdens_su}) is noteworthy; the two expressions are identical looking except for the interchange of $\mu$ and $\nu$ in the factors $(2\mu+2b+2)$ and $(-1)^\nu$ inside the double summation.
For large $N_1, N_2$ and fixed $N_1/N_2$, we obtain again (\ref{asymdens}). 

\begin{figure*}[t]
\centering
\includegraphics*[width=.98 \textwidth]{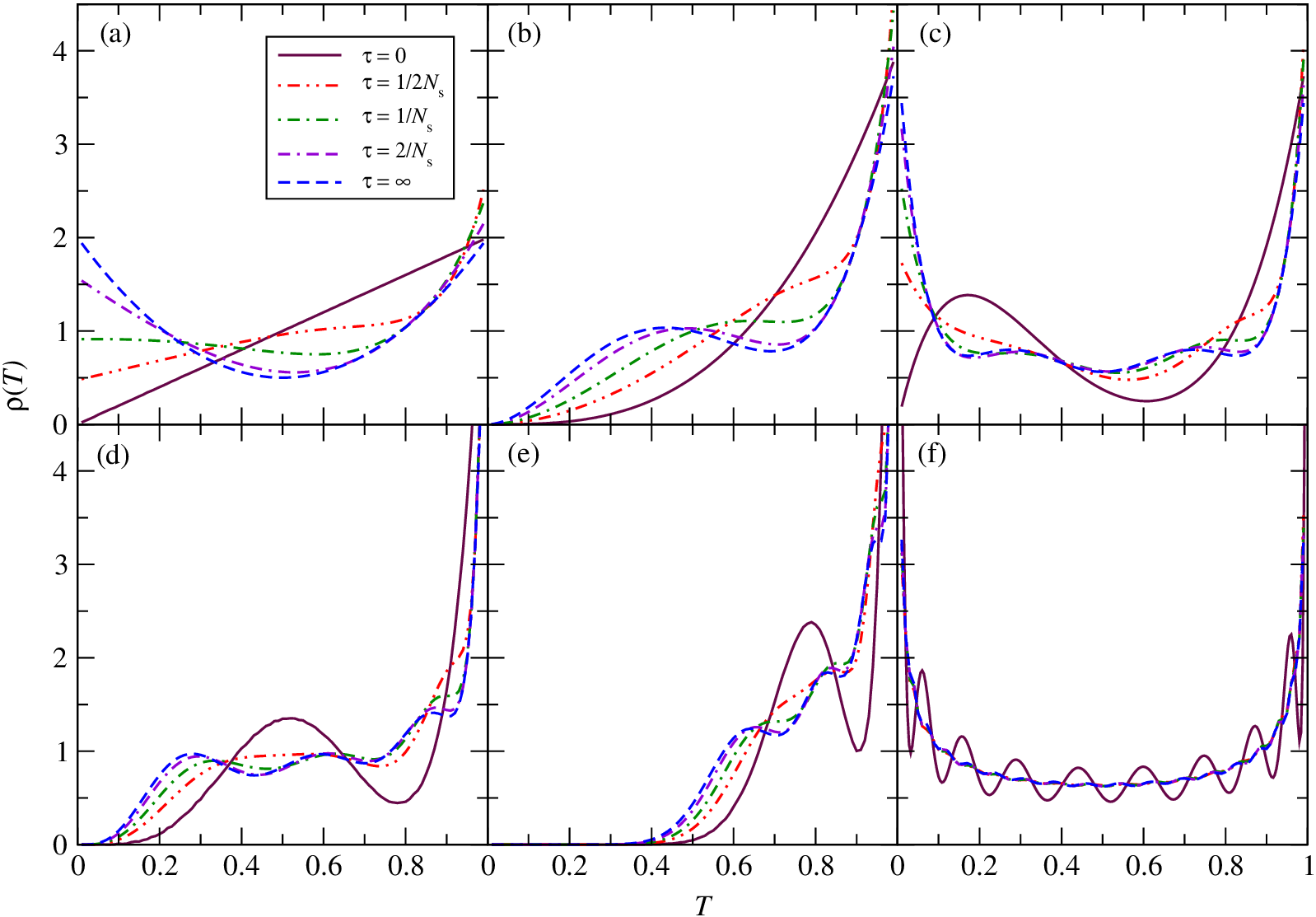}
\caption[]{The density of transmission eigenvalues $\rho(T)=\mathcal{R}_1(T)/N$ for SE-UE crossover at various $\tau$ for (a)$N_1=2,N_2=2$, (b)$N_1=2,N_2=4$, (c)$N_1=4,N_2=4$, (d)$N_1=4,N_2=8$, (e)$N_1=4,N_2=20$ and (f)$N_1=20,N_2=20$.}
\end{figure*}

Figure 2 shows the densities for different $N_1, N_2$ values as a function of $\tau$. Again, as in figure 1(e), the densities in figure 2(e) vanish below certain critical values. Also, as can be seen from figure 2(f), for large $N_1,N_2$ the density is again dominated by the first term corresponding to UE. However, in this case the oscillations do not die as rapidly as in the OE-UE crossover. 

We now give the results for the average conductance, average shot-noise power and variance of conductance for this crossover. The average conductance turns out to be
\begin{eqnarray}
\label{Gt-S}
\left<g\right>=\frac{N_1 N_2}{N_\mathrm{s}}+\frac{N_1 N_2}{N_\mathrm{s}(N_\mathrm{s}-1)}e^{-N_\mathrm{s}\tau}.
\end{eqnarray}
The average shot-noise power is obtained as
\begin{eqnarray}
\label{SNPt-S}
\fl
\left<p\right>=\frac{N_1^2N_2^2}{(N_\mathrm{s}-1)(N_\mathrm{s})(N_\mathrm{s}+1)}-\frac{N_1 N_2(N_1-N_2)^2}{(N_\mathrm{s}-2)(N_\mathrm{s}-1)(N_\mathrm{s})(N_\mathrm{s}+2)}e^{-N_\mathrm{s}\tau}\nonumber\\
\fl
+\frac{N_1 N_2(N_1-1)(N_2-1)}{(N_\mathrm{s}-3)(N_\mathrm{s}-2)(N_\mathrm{s}-1)(N_\mathrm{s})}e^{-2(N_\mathrm{s}-1)\tau}-\frac{N_1 N_2(N_1+1)(N_2+1)}{(N_\mathrm{s}-1)(N_\mathrm{s})(N_\mathrm{s}+1)(N_\mathrm{s}+2)}e^{-2(N_\mathrm{s}+1)\tau}.\nonumber
\\
\end{eqnarray}
The variance of conductance comes out as
\begin{eqnarray}
\label{varGt-S}
\fl
\mbox{var}(g)=\frac{N_1^2N_2^2}{(N_\mathrm{s}-1)(N_\mathrm{s})^2(N_\mathrm{s}+1)}+\frac{N_1 N_2(N_1-1)(N_2-1)}{(N_\mathrm{s}-3)(N_\mathrm{s}-2)(N_\mathrm{s}-1)(N_\mathrm{s})}e^{-2(N_\mathrm{s}-1)\tau}\nonumber\\
-\frac{N_1^2 N_2^2}{(N_\mathrm{s}-1)^2(N_s)^2}e^{-2N_\mathrm{s}\tau}+\frac{N_1 N_2(N_1+1)(N_2+1)}{(N_\mathrm{s}-1)(N_\mathrm{s})(N_\mathrm{s}+1)(N_\mathrm{s}+2)}e^{-2(N_\mathrm{s}+1)\tau}\nonumber\\
-\frac{2N_1 N_2(N_1-N_2)^2}{(N_\mathrm{s}-2)(N_\mathrm{s}-1)(N_\mathrm{s})^2(N_\mathrm{s}+2)}e^{-N_\mathrm{s}\tau}.
\end{eqnarray}
Again the exact expression for variance of shot-noise power is too complicated. We give the asymptotic result in section \ref{secUCF}.

Similar to OE-UE crossover the $\tau$-dependent terms lead to $\mathcal{O}(1/N_\mathrm{s})$ corrections in the respective leading terms in (\ref{Gt-S}) and (\ref{SNPt-S}) for large $N_1, N_2$. The sign of correction, however, is opposite to that in OE-UE crossover, hence the term weak antilocalization is used.

In $\tau=0$ limit (\ref{Gt-S})-(\ref{varGt-S}) agree with the known results (\ref{G_beta})-(\ref{VarG_beta}), provided two-fold degeneracy for each eigenvalue is properly considered. For that, along with $\beta=4$, we have to set $N_1\rightarrow N_1/2$, and $N_2\rightarrow N_2/2$ in  (\ref{G_beta})-(\ref{VarG_beta}), and then multiply the results for $\left<g\right>$, $\left<p\right>$ by 2 and the results for $\mbox{var}(g)$, $\mbox{var}(p)$ by 4 respectively. In the $\tau\rightarrow\infty$ limit the degeneracy no longer holds so the above results match with (\ref{G_beta})-(\ref{VarG_beta}) for $\beta=2$ without any modification.

We now point out an interesting observation for the OE-UE and SE-UE crossover results for $\left<g\right>$, $\left<p\right>$ and var($g$). One can obtain the results for SE-UE from OE-UE results and vice versa by simultaneously changing the sign of $N_1, N_2$ and $\tau$, and introducing an overall $-$ve sign for $\left<g\right>$, $\left<p\right>$. For var($g$) one does not need to introduce the overall $-$ve sign. 

%~~~~~~~~~~~~~~~~~~~~~ Section VI ~~~~~~~~~~~~~~~~~~~~~~~~
\section {Comparison with semiclassical results}
\label{secSC}

Weak time-reversal symmetry breaking condition is achieved when $\tau<<1/N_\mathrm{s}$, $\tau=1/N_\mathrm{s}$ being the natural crossover scale for the problem. In this case, with $N_1,N_2>>1$, we get from (\ref{Gt-O}), (\ref{SNPt-O}) and (\ref{Gt-S}), (\ref{SNPt-S}) the following results for the average conductance and average shot-noise power for the two transitions:
\begin{equation}
\label{G_asym}
\left<g\right>=\frac{N_1 N_2}{N_{\mathrm{s}}}\mp\frac{N_1 N_2}{N_{\mathrm{s}}^2(1+N_\mathrm{s}\tau)},
\end{equation}
\begin{equation}
\label{P_asym}
\left<p\right>=\frac{N_1^2N_2^2}{N_{\mathrm{s}}^3}\pm\frac{N_1 N_2(N_1-N_2)^2}{N_{\mathrm{s}}^4(1+N_\mathrm{s}\tau)}.
\end{equation}
In the above two equations the upper and lower signs correspond to the OE-UE and SE-UE transitions respectively. These results should be compared with the semiclassical predictions (for the OE-UE transition) for the average conductance and average shot-noise power when a weak perpendicular magnetic field is applied to the system \cite{BHMH, BC}, viz.
\begin{equation}
\left<g\right>=\frac{N_1 N_2}{N_{\mathrm{s}}}-\frac{N_1 N_2}{N_{\mathrm{s}}^2(1+\zeta)}+\mathcal{O}\left(\frac{1}{N_{\mathrm{s}}}\right),
\end{equation}
\begin{equation}
\left<p\right>=\frac{N_1^2N_2^2}{N_{\mathrm{s}}^3}+\frac{N_1 N_2(N_1-N_2)^2}{N_{\mathrm{s}}^4(1+\zeta)}+\mathcal{O}\left(\frac{1}{N_{\mathrm{s}}}\right),
\end{equation}
where $\zeta$ is proportional to the square of magnetic flux $\Phi$ through the system. (We have dropped here the extra factor of 2 appearing because of spin degeneracy on the RHS in \cite{BC}.) We know from the analysis done in \cite{KFP2} that $\tau$ is proportional to $\Phi^2$ for the quantities considered here.  Thus our results are consistent with semiclassical predictions. We believe that the semiclassical analysis for the SE-UE transition will also agree with the corresponding results in (\ref{G_asym}), (\ref{P_asym}). 

%~~~~~~~~~~~~~~~~~~~~~ Section VII ~~~~~~~~~~~~~~~~~~~~~~~~
\section {Universal fluctuations for conductance and shot-noise power}
\label{secUCF}

We have, from (\ref{varGt-O}) and (\ref{varGt-S}),
\begin{equation}
\label{varG_asym}
\mbox{var}(g)=\frac{N_1^2N_2^2}{N_{\mathrm{s}}^4}\big(1+e^{-2N_{\mathrm{s}}\tau}\big),
\end{equation}
for $N_1,N_2>>1$. This is valid for both the transitions. Equation (\ref{varG_asym}) leads to the phenomenon of universal conductance fluctuations \cite{MK,Bnkr,BM} in the $\tau=0$ and $\infty$ limits yielding the universal values 1/8 and 1/16 respectively. Identical result for var($g$) for $\beta=1$ and $\beta=4$ arise in the large $N_1, N_2$ limit because we have explicitly considered the Kramers degeneracy in SE. It indicates that the variance of conductance in the universal regime does not depend on the spin-rotation symmetry and is sensitive only to the time-reversal symmetry. Figure 3 shows comparison of the asymptotic result (\ref{varG_asym}) with the exact OE-UE and SE-UE crossover results.

\begin{figure*}[t]
\centering
\includegraphics*[width=.96\textwidth]{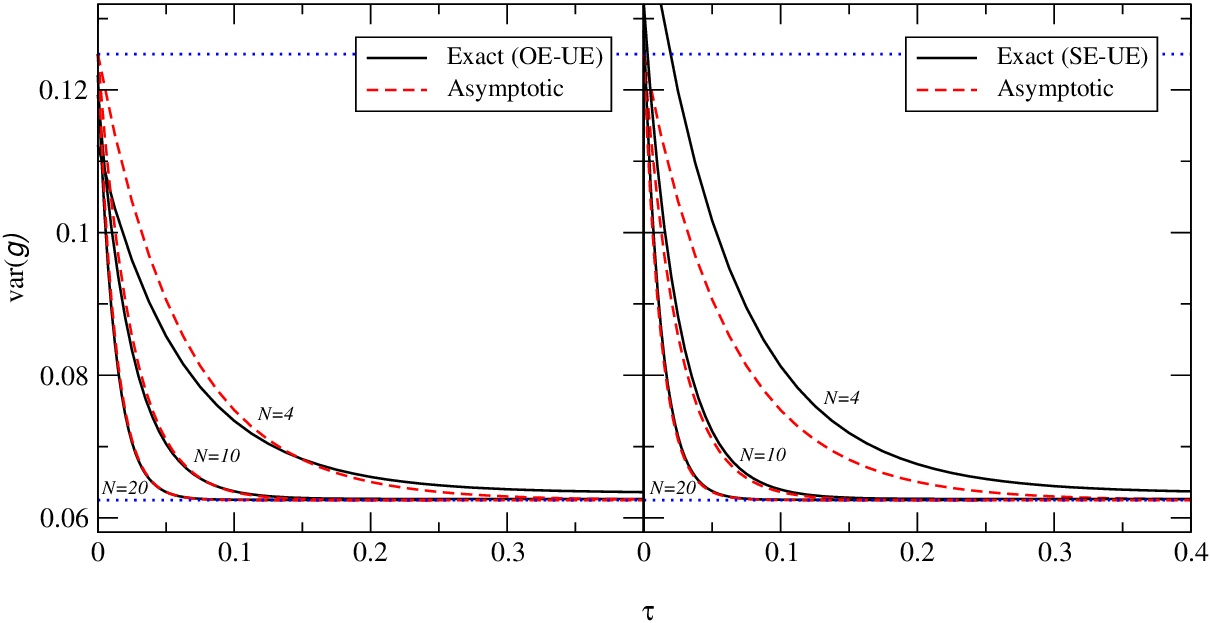}
\caption[]{The variance of conductance as a function of $\tau$ for various $N(=N_1=N_2)$ for the two transitions. The dotted horizontal lines represent the universal values 1/8 and 1/16.}
\end{figure*}

\begin{figure*}[t]
\centering
\includegraphics*[width=.96\textwidth]{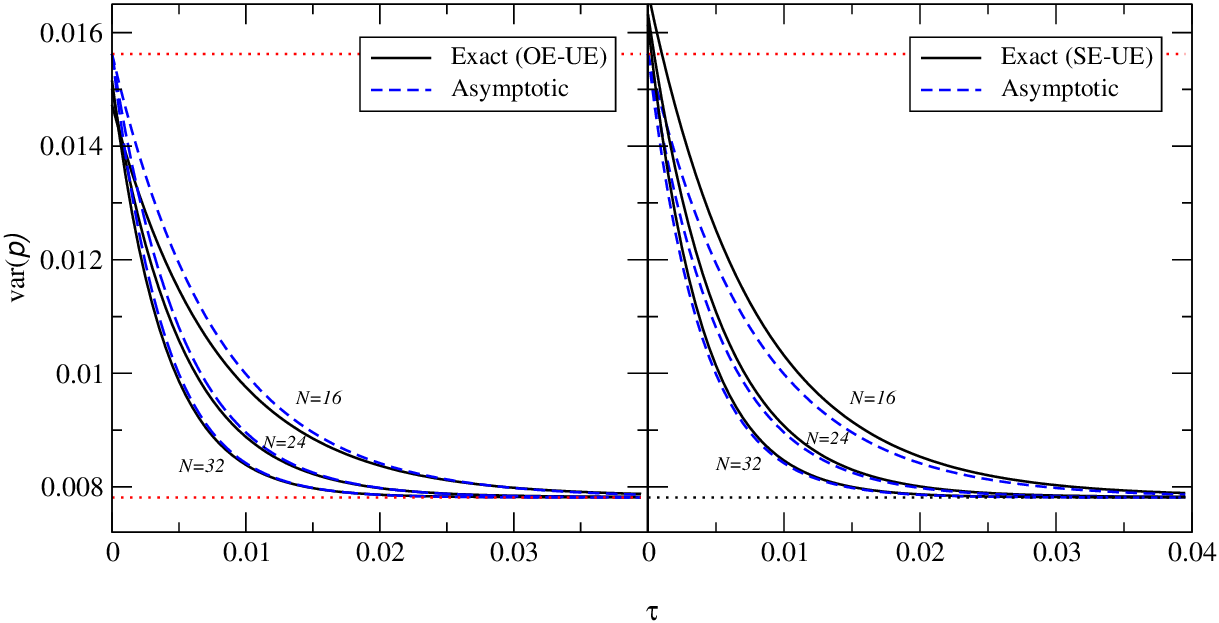}
\caption[]{The variance of shot-noise power as a function of $\tau$ for various $N(=N_1=N_2)$ for the two transitions. The dotted horizontal lines represent the universal values 1/64 and 1/128.}
\end{figure*}

We now consider the variance of shot-noise power. As mentioned earlier, the exact expression for var($p$) for the crossovers is quite complicated and lengthy. However for $N_1,N_2>>1$ it has a simple form. We find
\begin{equation}
\label{varP_asym}
\mbox{var}(p)=\frac{2N_1^4 N_2^4}{N_{\mathrm{s}}^8}\big(1+e^{-4N_{\mathrm{s}}\tau}\big)+\frac{N_1^2 N_2^2(N_1-N_2)^4}{N_{\mathrm{s}}^8}\big(1+e^{-2N_{\mathrm{s}}\tau}\big),
\end{equation}
valid as in (\ref{varG_asym}) for both the crossovers. In the $\tau=0$ and $\infty$ limits it leads to the universal values $1/64$ and $1/128$. The exact and asymptotic results for the variance of shot-noise power are compared in Figure 4.

An interesting quantity which involves $\left<g\right>, \left<p\right>$ and var($g$) collectively was suggested in \cite{SS}. The ratio $N_1 N_2\mbox{var}(g)/\left<g\right>\left<p\right>$ assumes the value $2/\beta$ for all $N_1, N_2$. It is natural to ask the behavior of this quantity for more general cases, for instance for the crossover ensembles. It is clear from the expressions of $\left<g\right>, \left<p\right>$ and var($g$) that in the crossover regime this quantity has, in general, a complicated dependence on $\tau$. However for $N_1,N_2>>1$, one obtains a rather simple form, similar to above expression for var($g$),
\begin{equation}
\label{q}
N_1 N_2\frac{\mbox{var}(g)}{\left<g\right>\left<p\right>}= 1+e^{-2N_{\mathrm{s}}\tau}.
\end{equation}
Note that taking into account the degeneracy for SE leads to identical expressions for the two crossovers. Figure 5 exhibits this quantity for several $N_1,N_2$ values. The OE-UE and SE-UE curves have been calculated using the exact results and the dotted curve represents the asymptotic results calculated using equation (\ref{q}). As can be seen this asymptotic result works nicely for $N_1$ and $N_2$ as low as 4.

\begin{figure}[ht]
\centering
\includegraphics*[width=.58\textwidth]{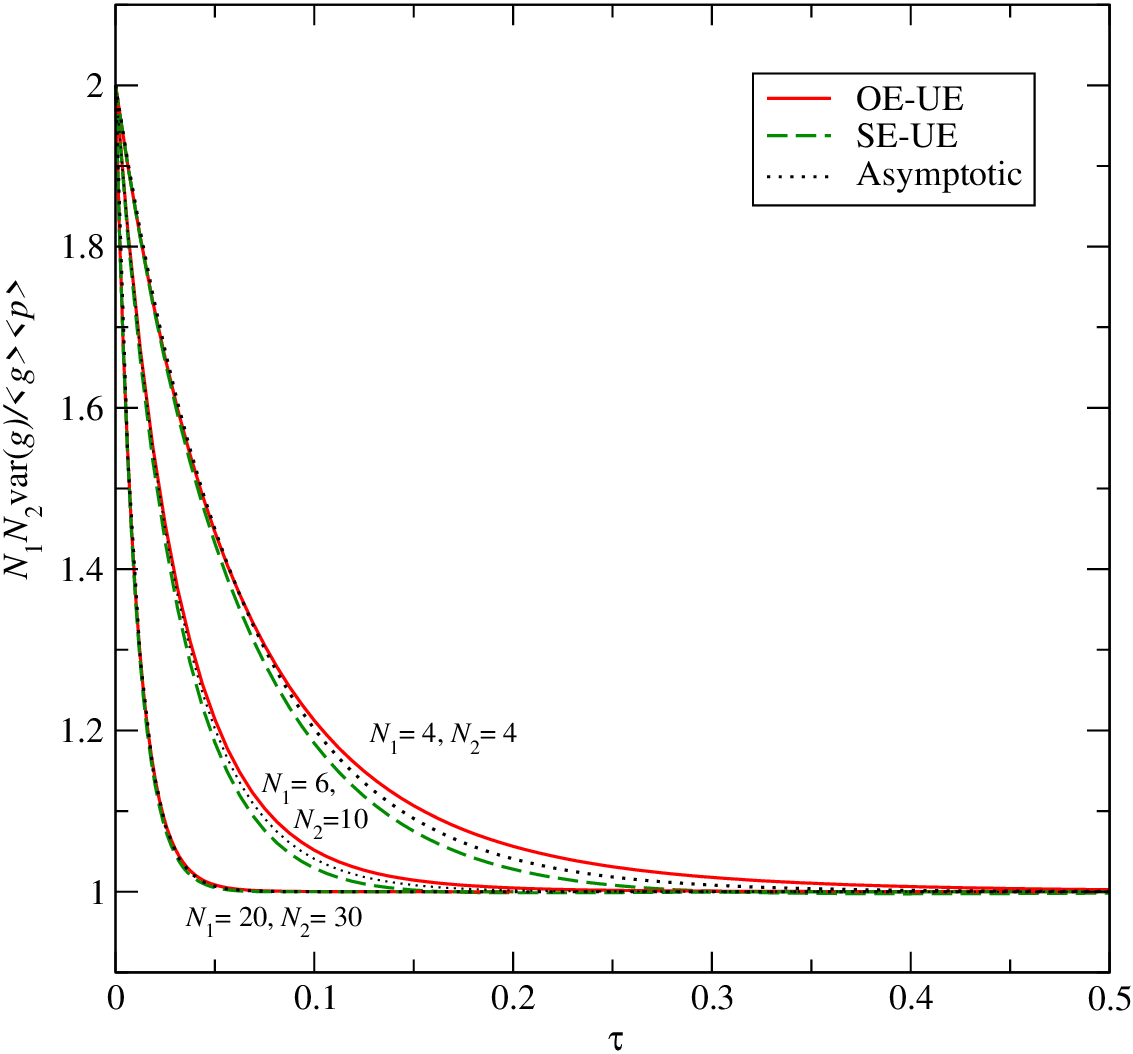}
\caption[]{$N_1 N_2\mbox{var}(g)/\left<g\right>\left<p\right>$ vs. $\tau$ for various $N_1, N_2$. The asymptotic curves have been drawn using (\ref{q}).} 
\end{figure}

%~~~~~~~~~~~~~~~~~~~~~ Section VIII ~~~~~~~~~~~~~~~~~~~~~~~~
\section {Conclusion}
\label{secC}

In conclusion, we have shown that the transmission eigenvalues in the OE-UE and SE-UE crossover regimes are described by the Jacobi crossover ensembles of random matrices. We have obtained exact results for the jpd of eigenvalues and the $n$-level correlation functions for these crossovers. By applying these results we have given expressions for averages and variances of conductance and shot-noise power as functions of Brownian motion parameter $\tau$. The relation of this parameter to the magnetic flux through the system has been discussed in detail by Frahm and Pichard \cite{KFP2}. 

The exact random-matrix expressions for the averages and variances of shot-noise power for arbitrary $N_1, N_2$ have become available only recently \cite{SS,SSW}. The $\tau$-dependent expressions derived here not only generalize the results for the averages and variances of conductance and shot-noise power to the crossover ensembles but also serve as alternative proofs for the $\beta=1,4$ and $\beta=2$ results in the $\tau\rightarrow0$ and $\tau\rightarrow\infty$ limits respectively.

Finally we have analyzed the results for large number of modes in the two leads. We have obtained the weak-localization corrections in the OE-UE case and the weak-antilocalization corrections in the SE-UE case for the average conductance and average shot-noise power. These are in agreement with the corresponding semiclassical results in the weak time-reversal regime of OE-UE crossover.

%~~~~~~~~~~~~~~~~~~~~~ Acknowledgement Section ~~~~~~~~~~~~~~~~~~~~~~~~
\ack

S. K. acknowledges CSIR India for financial assistance.

% ~~~~~~~~~~~~~~~~ Appendices ~~~~~~~~~~~~~~~~~
\appendix

%~~~~~~~~~~~~~~~~~~ Appendix A ~~~~~~~~~~~~~~~~~~~
\section{Jacobi polynomials} 
\setcounter{section}{1}

Jacobi polynomials $P_j^{\mu,\nu}(x)$, are defined with respect to the weight function $w_{\mu,\nu}(x)=(1-x)^\mu(1+x)^\nu$ in the interval $-1\le x \le 1$ as \cite{Sz}
\begin{equation}
\label{ortho}
\int_{-1}^{1}w_{\mu,\nu}(x)P_j^{\mu,\nu}(x)P_k^{\mu,\nu}(x)dx=h_j^{\mu,\nu}\delta_{jk}.
\end{equation}
Here $h_j^{\mu,\nu}$ is the normalization and is given by
\begin{equation}
h_j^{\mu,\nu}=\frac{2^{\mu+\nu+1}}{(2j+\mu+\nu+1)}\frac{\Gamma(j+\mu+1)\Gamma(j+\nu+1)}{\Gamma(j+1)\Gamma(j+\mu+\nu+1)}.
\end{equation}
In the appendices that follow we use the notation
\begin{equation}
 \mathcal{P}_n(x)\equiv P_n^{0,2b+1}(x)
\end{equation}
for compactness. $\mathcal{P}_n(x)$ should not be confused with jpd in (\ref{PeqT}). 
Similarly we use
\begin{equation}
\label{norm}
h_j\equiv h_j^{0,2b+1}=\frac{2^{2b+1}}{(j+b+1)}.
\end{equation}
In the main text, however, we stick to the original notation.

The three-term recurrence relation \cite{Sz} satisfied by Jacobi polynomials simplifies to the following for $\mathcal{P}_n(x)$:
\begin{eqnarray}
\nonumber
x\mathcal{P}_n(x)\!\!&=&\!\!\frac{(2n+2)(n+2b+2)}{(2n+2b+2)(2n+2b+3)}\mathcal{P}_{n+1}(x)\\
\nonumber
&+&\!\!\frac{(2b+1)^2}{(2n+2b+1)(2n+2b+3)}\mathcal{P}_{n}(x)\\
&+&\!\!\frac{(2n)(n+2b+1)}{(2n+2b+1)(2n+2b+2)}\mathcal{P}_{n-1}(x).
\end{eqnarray}
This, along with the orthogonality relation (\ref{ortho}), is used in calculating the moments (\ref{mx})-(\ref{mxy}).

%~~~~~~~~~~~~~~~~~~ Appendix B ~~~~~~~~~~~~~~~~~~~
\section{One-body operator \texorpdfstring{$\boldsymbol{O}_x$}{Ox} and the antisymmetric function \texorpdfstring{$F_{jk}$}{Fjk}} 
\setcounter{section}{2}

The one-body operator introduced in \cite{SKP}, which is used in finding the expressions appropriate to the crossover from $\tau=0$ results, is given by
\begin{equation}
\boldsymbol{O}_x=\frac{\sqrt{w_{0,2b+1}(x)}}{w(x)}e^{-\mathcal{H}_x\tau}\frac{w(x)}{\sqrt{w_{0,2b+1}(x)}}.
\end{equation}
Here $w(x)$ is same as that in (\ref{Ptrans}) and, as explained earlier, equals $w_{0,b}(x)$ and $w_{0,b+1}(x)$ for OE and SE respectively. Also $w_{0,2b+1}(x)$ is the weight function corresponding to the UE. Since $\mathcal{H}_x$ has the eigenfunctions $w_{0,b+1/2}(x)\mathcal{P}_n(x)$ with eigenvalues $\boldsymbol{\varepsilon}_n$, the eigenfunctions of $\boldsymbol{O}_x$ are $(w_{0,2b+1}(x)/w(x))\mathcal{P}_n(x)$ with eigenvalues $e^{-\boldsymbol{\varepsilon}_n\tau}$. Similarly the eigenfunctions of the operator $(\boldsymbol{O}_x^\dag)^{-1}$ are $w(x)\mathcal{P}_n(x)$ with eigenvalues $e^{\boldsymbol{\varepsilon}_n\tau}$.

The antisymmetric function $F_{j,k}^{(\tau)}$ in (\ref{Ptrans}) can be obtained from its $\tau=0$ counterpart by using the operator $\boldsymbol{O}$. For SE-UE crossover and also for the even $N$ case of OE-UE crossover it equals $\mathcal{G}^{(\tau)}(x_j,x_k)$ with $j,k=1,2,...,N$. In terms of the functions dual to skew-orthogonal polynomials (see appendix C) it is given by
\begin{eqnarray}
\label{Gxyt}
\mathcal{G}^{(\tau)}(x,y)=\sum_{\mu=0}^{\infty}[\psi_{2\mu}^{(\tau)}(x)\psi_{2\mu+1}^{(\tau)}(y)-\psi_{2\mu+1}^{(\tau)}(x)\psi_{2\mu}^{(\tau)}(y)].
\end{eqnarray} 
$\mathcal{G}^{(\tau)}(x_j,x_k)$ is obtained by application of $\boldsymbol{O}_{x_j}$ and $\boldsymbol{O}_{x_k}$ on its $\tau=0$ counterpart, viz.,
\begin{equation}
\mathcal{G}^{(\tau)}(x_j,x_k)=\boldsymbol{O}_{x_j}\boldsymbol{O}_{x_k}\mathcal{G}^{(0)}(x_j,x_k).
\end{equation}
 For the $N$ odd case of OE-UE crossover we also need to introduce
\begin{equation}
\label{omg}
\omega^{(\tau)}(x)=\frac{w_{0,b+1}(x)}{{2^b}}\sum_{\mu=0}^{\infty}(-1)^\mu e^{-\varepsilon_\mu\tau}\mathcal{P}_\mu(x),
\end{equation}
then the expression for jpd of eigenvalues (\ref{Ptrans}) holds with the addition $F_{j,N+1}^{(\tau)}=-F_{N+1,j}^{(\tau)}=\frac{1}{2}\omega^{(\tau)}(x_j)(1-\delta_{j,N+1})$. 

Note that for the odd $N$ case of OE $\omega^{(0)}(x)=1$. To obtain $\omega^{(\tau)}(x)$ from this we use $1\equiv\int_{-1}^{1}\delta(z-x)dz$ and the expansion
\begin{equation}
\delta(z-x)=w_{0,b}(z)w_{0,b+1}(x)\sum_{\mu=0}^{\infty}\frac{1}{h_\mu}\mathcal{P}_\mu(z)\mathcal{P}_\mu(x).
\end{equation} 
Integration over $z$ then leads to (\ref{omg}) with $\tau=0$. Operation of $\boldsymbol{O}_x$ on this finally leads to the expression valid for arbitrary $\tau$ (see (\ref{OmgEvol})).

%~~~~~~~~~~~~~~~~~~~ Appendix C ~~~~~~~~~~~~~~~~~~~
\section{Skew-orthogonal polynomials} 
\setcounter{section}{3}

The weighted skew-orthogonal polynomials $\phi_j^{(\tau)}(x)$ and the dual functions $\psi_j^{(\tau)}(x)$ satisfy the following skew orthogonality relation:
\begin{equation}
\label{skew}
\int_{-1}^{1}\phi_j^{(\tau)}(x)\psi_k^{(\tau)}(x)dx=Z_{jk},
\end{equation}
where $Z_{jk}=-Z_{kj}$ equals 1 for $k=j+1$ with $j$ even, $-1$ for $k=j-1$ with $j$ odd, and 0 for $|j-k|\neq 1$. Here
\begin{equation}
 \psi_j^{(\tau)}(x)=\int_{-1}^{1}\mathcal{G}^{(\tau)}(x,y) \phi_j^{(\tau)}(y)dy.
\end{equation}
For the odd $N$ case of OE-UE crossover in addition to the above skew orthogonality relation we also have the extra condition, similar to the $\tau=0$ case,\cite{PG2000,GP}
\begin{equation}
\int_{-1}^{1}\omega^{(\tau)}(x)\phi_j^{(\tau)}(x)dx=\delta_{j,N-1}
\end{equation}
because of the presence of the unpaired $\phi_{N-1}^{(\tau)}(x)$.

These $\tau$-dependent functions are obtained from their $\tau=0$ counterparts\cite{PG2000,GP} by application of the operators $\boldsymbol{O}_x$ and $(\boldsymbol{O}_x^\dag)^{-1}$ as
\begin{equation}
\label{PhiEvol}
\phi_j^{(\tau)}(x)=(\boldsymbol{O}_x^\dag)^{-1}\phi_j^{(0)}(x),
\end{equation}
\begin{equation}
\label{PsiEvol}
\psi_j^{(\tau)}(x)=\boldsymbol{O}_x\psi_j^{(0)}(x),
\end{equation}
\begin{equation}
\label{OmgEvol}
\omega^{(\tau)}(x)=\boldsymbol{O}_x\,\omega^{(0)}(x).
\end{equation} 

% ~~~~~~~~~~~~~~ Appendix D ~~~~~~~~~~~~~~~~~~~
\section{Two point kernels and the \texorpdfstring{$n$}{n}-level correlation function} 
\setcounter{section}{4}

Let 
\begin{equation}
c=\cases{
  0, & $N$ even,\\
  1, & $N$ odd.
        }
 \end{equation} 
The two point kernels are then defined in the following manner:
\begin{eqnarray}
\nonumber
\label{Sxyt}
S_N^{(\tau)}(x,y)&=&\!\!\!\sum_{\mu=0}^{(\frac{N-c}{2})-1}[\phi_{2\mu}^{(\tau)}(x)\psi_{2\mu+1}^{(\tau)}(y)-\phi_{2\mu+1}^{(\tau)}(x)\psi_{2\mu}^{(\tau)}(y)]\\
&+&c\;\phi_{N-1}^{(\tau)}(x)\omega^{(\tau)}(y),
\end{eqnarray}
\begin{eqnarray}
\label{Axyt}
\nonumber
A_N^{(\tau)}(x,y)&=&\!\!\!\sum_{\mu=0}^{(\frac{N-c}{2})-1}[\phi_{2\mu+1}^{(\tau)}(x)\phi_{2\mu}^{(\tau)}(y)-\phi_{2\mu}^{(\tau)}(x)\phi_{2\mu+1}^{(\tau)}(y)],\\&&
\end{eqnarray}
\begin{eqnarray}
\label{Bxyt}
\nonumber
B_N^{(\tau)}(x,y)&=&\!\!\!\sum_{\mu=(\frac{N-c}{2})}^{\infty}[\psi_{2\mu+1}^{(\tau)}(x)\psi_{2\mu}^{(\tau)}(y)-\psi_{2\mu}^{(\tau)}(x)\psi_{2\mu+1}^{(\tau)}(y)]\\
&+&c\,[\psi_{N-1}^{(\tau)}(x)\omega^{(\tau)}(y)-\psi_{N-1}^{(\tau)}(y)\omega^{(\tau)}(x)].
\end{eqnarray} 
Note that for SE-UE crossover $N$ is even only. These kernels can be obtained by applying the operators $(\boldsymbol{O}_x^\dag)^{-1}$ and $\boldsymbol{O}_x$ on the kernels for $\tau=0$ using (\ref{PhiEvol}), (\ref{PsiEvol}) and (\ref{OmgEvol}). Now the $n$-level correlation function can be written in terms of a quaternion determinant as
\begin{equation}
\label{Qdet}
R_n^{(\tau)}(x_1,...,x_n)=\mbox{Qdet}\left(
\begin{array}{cc}
S_N^{(\tau)}(x_j,x_k) & A_N^{(\tau)}(x_j,x_k)\\
B_N^{(\tau)}(x_j,x_k) & S_N^{(\tau)}(x_k,x_j)
\end{array}\right)_{j,k=1,...,n}\!\!\!.
\end{equation} 

% ~~~~~~~~~~~~~~~~ Appendix E ~~~~~~~~~~~~~~~~~
\section{Skew-orthogonal polynomials for OE-UE crossover} 
\setcounter{section}{5}

In this case we have $\mathcal{G}^{(0)}(x,y)=\mbox{sgn}(x-y)/2$. The weighted skew-orthogonal polynomials $\phi_j^{(0)}(x)$ and the dual functions $\psi_j^{(0)}(x)$ for the initial weight function $w_{0,b}(x)$ (corresponding to OE) can be compactly written down in terms of the Jacobi polynomials $P_j^{1,2b+1}(x)$ using the results given in \cite{PG2000,GP}. The eigenfunctions of $\boldsymbol{O}_x$ however involve $\mathcal{P}_j(x)$. We therefore need to expand $\phi_j^{(0)}(x)$ and $\psi_j^{(0)}(x)$ in terms of $\mathcal{P}_j(x)$ prior to the applications of operators to calculate the corresponding expressions for the transition. Such expansions are given in \cite{Sz}:
\begin{equation}
\label{Sz1}
P_n^{1,2b+1}(x)=\frac{2}{(n+2b+2)}\sum_{\mu=0}^n (\mu+b+1)\mathcal{P}_\mu(x),
\end{equation}
\begin{equation}
\label{Sz2}
(1-x)P_n^{1,2b+1}(x)=\frac{2(n+1)}{(2n+2b+3)}\left[\mathcal{P}_n(x)-\mathcal{P}_{n+1}(x)\right].
\end{equation}
Thus, using (\ref{PhiEvol}) and (\ref{PsiEvol}) we find, for even $N$,
\begin{equation}
\label{ephie}
\phi_{2m}^{(\tau)}(x)=2^{b+1/2}w_{0,b}(x)\sum_{\mu=0}^{2m}\frac{e^{\boldsymbol{\varepsilon}_\mu\tau}}{h_\mu}\mathcal{P}_\mu(x),~~~~~~~~~~
\end{equation}
\begin{equation}
\label{epsie}
\psi_{2m}^{(\tau)}(x)=-\frac{w_{0,b+1}(x)}{2^{b+1/2}}\!\!\sum_{\mu=2m+1}^\infty(-1)^\mu e^{-\boldsymbol{\varepsilon}_\mu\tau}\mathcal{P}_{\mu}(x),
\end{equation}
\begin{eqnarray}
\label{ephio}
\phi_{2m+1}^{(\tau)}(x)=-2^{b+1/2}w_{0,b}(x)\Big[\frac{e^{\boldsymbol{\varepsilon}_{2m}\tau}}{h_{2m}}\mathcal{P}_{2m}(x)+\frac{e^{\boldsymbol{\varepsilon}_{2m+1}\tau}}{h_{2m+1}}\mathcal{P}_{2m+1}(x)\Big],
\end{eqnarray}
\begin{eqnarray}
\label{epsio}
\psi_{2m+1}^{(\tau)}(x)=\frac{w_{0,b+1}(x)}{2^{b+1/2}}\Big[e^{-\boldsymbol{\varepsilon}_{2m}\tau}\mathcal{P}_{2m}(x)-e^{-\boldsymbol{\varepsilon}_{2m+1}\tau}\mathcal{P}_{2m+1}(x)\Big].
\end{eqnarray}
for $m=0,1,2,...$ .
When $N$ is odd, we have for $m=0,1,...,(N-1)/2-1$,
\begin{equation}
\label{ophie}
\phi_{2m}^{(\tau)}(x)=2^{b+1/2}w_{0,b}(x)\sum_{\mu=0}^{2m+1}\frac{e^{\boldsymbol{\varepsilon}_\mu\tau}}{h_\mu}\mathcal{P}_\mu(x),
\end{equation}
\begin{equation}
\label{opsie}
\psi_{2m}^{(\tau)}(x)=-\frac{w_{0,b+1}(x)}{2^{b+1/2}}\sum_{\mu=0}^{2m+1}(-1)^{\mu}e^{-\boldsymbol{\varepsilon}_\mu\tau}\mathcal{P}_{\mu}(x),
\end{equation}
\begin{eqnarray}
\label{ophio}
\phi_{2m+1}^{(\tau)}(x)=-2^{b+1/2}w_{0,b}(x)\Big[\frac{e^{\boldsymbol{\varepsilon}_{2m+1}\tau}}{h_{2m+1}}\mathcal{P}_{2m+1}(x)+\frac{e^{\boldsymbol{\varepsilon}_{2m+2}\tau}}{h_{2m+2}}\mathcal{P}_{2m+2}(x)\Big],
\end{eqnarray}
\begin{eqnarray}
\label{opsio}
\psi_{2m+1}^{(\tau)}(x)=\frac{w_{0,b+1}(x)}{2^{b+1/2}}\Big[e^{-\boldsymbol{\varepsilon}_{2m+1}\tau}\mathcal{P}_{2m+1}(x)-e^{-\boldsymbol{\varepsilon}_{2m+2}\tau}\mathcal{P}_{2m+2}(x)\Big].
\end{eqnarray}
The unpaired $\phi_{N-1}^{(\tau)}(x)$ turns out to be
\begin{equation}
\phi_{N-1}^{(\tau)}(x)=2^b w_{0,b}(x)\sum_{\mu=0}^{N-1}\frac{e^{\boldsymbol{\varepsilon}_{\mu}\tau}}{h_\mu}\mathcal{P}_{\mu}(x),
\end{equation} 
along with the $\psi$ function
\begin{equation}
\psi_{N-1}^{(\tau)}(x)=-\frac{w_{0,b+1}(x)}{2^{b+1}}\sum_{\mu=N}^{\infty}(-1)^{\mu} e^{-\boldsymbol{\varepsilon}_{\mu}\tau}\mathcal{P}_{\mu}(x).
\end{equation}
The $\tau=0$ expressions, from which the above results follow, are contained in the above equations.

%~~~~~~~~~~~~~~~~ Appendix F ~~~~~~~~~~~~~~~~
\section{Skew-orthogonal polynomials for SE-UE crossover} 
\setcounter{section}{6}

The skew-orthogonal polynomials for SE-UE crossover again satisfy the relation (\ref{skew}), however unlike OE-UE crossover here we have $\mathcal{G}^{(0)}(x,y)=-\delta'(x-y)$. As explained in section \ref{secSU}, in this case the $\tau=0$ polynomials correspond to the initial weight function $w_{0,b+1}(x)$. Since this weight function does not vanish at $x=1$, so extra care has to be taken in finding the appropriate skew-orthogonal polynomials which satisfy (\ref{skew}). To calculate the $\psi_j^{(0)}(x)$ using $\int_{-1}^{1}\mathcal{G}^{(0)}(x,y)\phi_j^{(0)}(y)dy$, one has to take into account the contribution from boundary term also. Also the following relations turn out to be useful in arriving at $\tau=0$ expansions of $\phi$ and $\psi$ involving $\mathcal{P}_j(x)$:
\begin{equation}
\frac{P_n^{-1,2b+1}(x)}{(1-x)}=-\frac{1}{n}\sum_{\mu=1}^n(\mu+b)\mathcal{P}_{\mu-1}(x)~~~~(n\ge1),
\end{equation}
\begin{equation}
P_n^{-1,2b+1}(x)=\frac{(n+2b+1)}{(2n+2b+1)}\left[\mathcal{P}_n(x)-\mathcal{P}_{n-1}(x)\right]~~(n\ge0),
\end{equation}
along with $P_{-1}^{\mu,\nu}(x)\equiv0$. It is to be noted here that the polynomials $P_\mu^{-1,2b+1}(x)$ form orthogonal set for $\mu\ge 1$ only. 
These relations follow from the results given in \cite{Sz}. 

Once the $\tau=0$ results are known the results for arbitrary $\tau$ can be written down easily using the operators $\boldsymbol{O}_x$ and $(\boldsymbol{O}_x^\dag)^{-1}$. The final results for the $\phi$ and $\psi$ functions for this crossover read
\begin{equation}
\label{phie}
\phi_{2m}^{(\tau)}(x)=-\frac{w_{0,b+1}(x)}{2^{b+1/2}}\sum_{\mu=0}^{2m}(-1)^{\mu}e^{\boldsymbol{\varepsilon}_\mu\tau}\mathcal{P}_{\mu}(x),
\end{equation}
\begin{equation}
\label{psie}
\psi_{2m}^{(\tau)}(x)=-2^{b+1/2}w_{0,b}(x)\!\sum_{\mu=2m+1}^{\infty}\frac{e^{-\boldsymbol{\varepsilon}_\mu\tau}}{h_\mu}\mathcal{P}_{\mu}(x),
\end{equation}
\begin{eqnarray}
\label{phio}
\phi_{2m+1}^{(\tau)}(x)\!\!&=&\!\!-\frac{w_{0,b+1}(x)}{2^{b+1/2}}\Big[e^{\boldsymbol{\varepsilon}_{2m}\tau}\mathcal{P}_{2m}(x)-e^{\boldsymbol{\varepsilon}_{2m+1}\tau}\mathcal{P}_{2m+1}(x)\Big],
\end{eqnarray}
\begin{eqnarray}
\label{psio}
\psi_{2m+1}^{(\tau)}(x)\!\!&=&\!\!-2^{b+1/2}w_{0,b}(x)\Big[\frac{e^{-\boldsymbol{\varepsilon}_{2m}\tau}}{h_{2m}}\mathcal{P}_{2m}(x)+\frac{e^{-\boldsymbol{\varepsilon}_{2m+1}\tau}}{h_{2m+1}}\mathcal{P}_{2m+1}(x)\Big].
\end{eqnarray}

%~~~~~~~~~~~~~~~~~ Appendix G ~~~~~~~~~~~~~~~~~~
\section{Two-point kernels for OE-UE crossover} 
\setcounter{section}{7}

We substitute the expressions of $\phi_j^{(\tau)}(x)$ and $\psi_j^{(\tau)}(x)$ in (\ref{Sxyt})-(\ref{Bxyt}) and (\ref{Gxyt}). By manipulation of the summations we obtain the following expressions for the kernels which are valid for both even and odd $N$:
\begin{eqnarray}
\label{Ssum_ou}
\nonumber
\fl
\!\!S_N^{(\tau)}(x,y)=w_{0,b}(x)w_{0,b+1}(y)\Big[\!\!\sum_{\mu=0}^{N-1}\frac{1}{h_\mu}\mathcal{P}_\mu(x)\mathcal{P}_\mu(y)
-\!\!\sum_{\mu=0}^{N-1}\sum_{\nu=N}^{\infty}\frac{(-1)^{\nu+N} e^{-(\varepsilon_\nu-\varepsilon_\mu)\tau}}{h_{\mu}} \mathcal{P}_{\mu}(x)\mathcal{P}_{\nu}(y)\Big]\!,\\
\end{eqnarray}
\begin{eqnarray}
\label{Asum_ou}
\nonumber
\fl
\!\!A_N^{(\tau)}(x,y)=-w_{0,b}(x)w_{0,b}(y)\sum_{\mu=1}^{N-1}\sum_{\nu=0}^{\mu-1}
\frac{2^{2b+1}e^{(\varepsilon_\mu+\varepsilon_\nu)\tau}}{h_\mu h_\nu}\Big[\mathcal{P}_{\mu}(x)\mathcal{P}_{\nu}(y)-\mathcal{P}_{\nu}(x)\mathcal{P}_{\mu}(y)\Big],\\
\end{eqnarray}
\begin{eqnarray}
\label{Bsum_ou}
\fl
\!\!B_N^{(\tau)}(x,y)=-w_{0,b+1}(x)w_{0,b+1}(y)\!\!\sum_{\mu=N}^{\infty}\!\sum_{\nu=\mu+1}^{\infty}
\frac{(-1)^{\mu+\nu}e^{-(\varepsilon_\mu+\varepsilon_\nu)\tau}}{2^{2b+1}}\Big[\mathcal{P}_{\mu}(x)\mathcal{P}_{\nu}(y)-\mathcal{P}_{\nu}(x)\mathcal{P}_{\mu}(y)\Big],\nonumber\\
\hspace{-4cm}~
\end{eqnarray}
\begin{eqnarray}
\label{Gsum_ou}
\fl
\nonumber
\mathcal{G}^{(\tau)}(x,y)=                                                             w_{0,b+1}(x)w_{0,b+1}(y)\sum_{\mu=0}^{\infty}\sum_{\nu=\mu+1}^{\infty}
\frac{(-1)^{\mu+\nu}e^{-(\varepsilon_\mu+\varepsilon_\nu)\tau}}{2^{2b+1}}\Big[\mathcal{P}_{\mu}(x)\mathcal{P}_{\nu}(y)-\mathcal{P}_{\nu}(x)\mathcal{P}_{\mu}(y)\Big]\\
\nonumber
\fl
~~~~~~~~~~~~=                                                             -w_{0,b+1}(x)w_{0,b+1}(y)\sum_{\mu=1}^{\infty}\sum_{\nu=0}^{\mu-1}
\frac{(-1)^{\mu+\nu}e^{-(\varepsilon_\mu+\varepsilon_\nu)\tau}}{2^{2b+1}}\Big[\mathcal{P}_{\mu}(x)\mathcal{P}_{\nu}(y)-\mathcal{P}_{\nu}(x)\mathcal{P}_{\mu}(y)\Big].\\
\end{eqnarray}
The first expansion of $\mathcal{G}^{(\tau)}(x,y)$ in (\ref{Gsum_ou}) is obtained by using the $\phi$ and $\psi$ belonging to the even-$N$ case in (\ref{Gxyt}), whereas the second one follows by using $\phi$ and $\psi$ belonging to the odd-$N$ case. As expected, these two representations for $\mathcal{G}^{(\tau)}(x,y)$ are equivalent and follow from each other by changing the order of summations.

%~~~~~~~~~~~~~~~~~ Appendix H ~~~~~~~~~~~~~~~~~~~
\section{Two-point kernels for SE-UE crossover} 
\setcounter{section}{8}

For the SE-UE transition we obtain
\begin{eqnarray}
\label{Ssum_su}
\nonumber
\fl
\!\!S_N^{(\tau)}(x,y)=w_{0,b+1}(x)w_{0,b}(y)\Big[\sum_{\mu=0}^{N-1}\frac{1}{h_\mu}\mathcal{P}_\mu(x)\mathcal{P}_\mu(y)
-\sum_{\mu=0}^{N-1}\sum_{\nu=N}^{\infty}\frac{(-1)^{\mu} e^{-(\varepsilon_\nu-\varepsilon_\mu)\tau}}{h_{\nu}} \mathcal{P}_{\mu}(x)\mathcal{P}_{\nu}(y)\Big],\\
\end{eqnarray}
\begin{eqnarray}
\label{Asum_su}
\nonumber
\fl
A_N^{(\tau)}(x,y)=w_{0,b+1}(x)w_{0,b+1}(y)\sum_{\mu=1}^{N-1}\sum_{\nu=0}^{\mu-1}
\frac{(-1)^{\mu+\nu}e^{(\varepsilon_\mu+\varepsilon_\nu)\tau}}{2^{2b+1}}\Big[\mathcal{P}_{\mu}(x)\mathcal{P}_{\nu}(y)-\mathcal{P}_{\nu}(x)\mathcal{P}_{\mu}(y)\Big],\\
\end{eqnarray}
\begin{eqnarray}
\label{Bsum_su}
\nonumber
\fl
B_N^{(\tau)}(x,y)=w_{0,b}(x)w_{0,b}(y)\sum_{\mu=N}^{\infty}\sum_{\nu=\mu+1}^{\infty}\frac{2^{2b+1}e^{-(\varepsilon_\mu+\varepsilon_\nu)\tau}}{h_\mu h_\nu}\Big[\mathcal{P}_{\mu}(x)\mathcal{P}_{\nu}(y)-\mathcal{P}_{\nu}(x)\mathcal{P}_{\mu}(y)\Big],\\
\end{eqnarray}
\begin{eqnarray}
\label{Gsum_su}
\nonumber
\fl
\mathcal{G}^{(\tau)}(x,y)=-w_{0,b}(x)w_{0,b}(y)\sum_{\mu=0}^{\infty}\sum_{\nu=\mu+1}^{\infty}\frac{2^{2b+1}e^{-(\varepsilon_\mu+\varepsilon_\nu)\tau}}{h_\mu h_\nu}\Big[\mathcal{P}_{\mu}(x)\mathcal{P}_{\nu}(y)-\mathcal{P}_{\nu}(x)\mathcal{P}_{\mu}(y)\Big]\\
\nonumber
\fl
~~~~~~~~~~~~=w_{0,b}(x)w_{0,b}(y)\sum_{\mu=1}^{\infty}\sum_{\nu=0}^{\mu-1}\frac{2^{2b+1}e^{-(\varepsilon_\mu+\varepsilon_\nu)\tau}}{h_\mu h_\nu}\Big[\mathcal{P}_{\mu}(x)\mathcal{P}_{\nu}(y)-\mathcal{P}_{\nu}(x)\mathcal{P}_{\mu}(y)\Big].\\
\end{eqnarray}
In (\ref{Gsum_su}) the first expansion follows from the substitution of $\phi$ and $\psi$ for SE-UE crossover in (\ref{Gxyt}). We have given the second expansion as an analogy to the corresponding expansion in (\ref{Gsum_ou}) by changing the order of summation in the first one.

% ~~~~~~~~~~~~~~~~~~

\end{document}